\documentclass[preprint, pre, amsmath, amssymb]{revtex4}

\usepackage{graphicx}
\usepackage{bm}

\begin{document}

\title{Action potential restitution and hysteresis in a reaction-diffusion system with pacing rate dependent excitation threshold}

\author{J. M. Starobin}
\email{jmstarob@uncg.edu}
\affiliation{University of North Carolina at Greensboro, Greensboro, NC} 
\affiliation{Mediwave Star Technology, Inc., Greensboro, NC}

\author{C. P. Danford}
\author{V. Varadarajan}
\author{A. J. Starobin}
\author{V. N. Polotski}

\affiliation{Mediwave Star Technology, Inc., Greensboro, NC}

\date{\today}

\begin{abstract}
We have demonstrated that rate dependent restitution and action potential duration-refractory period hysteresis can be reproduced in a one-dimensional two-variable Chernyak-Starobin-Cohen reaction-diffusion medium with variable excitation threshold. We show that restitution and hysteresis depend on the relationship between pacing period and steady state excitation threshold and also on the rate of excitation threshold adaptation after an abrupt change in pacing period. It was also observed that the onset of action potential duration alternans is determined by the minimal stable wavefront speed, which could be approximated by the analytical critical speed of a stable solitary pulse. This approximation was suitably accurate regardless of the adaptation constant of excitation threshold, its dependence on pacing interval, or magnitude of the slopes of restitution curves.
\end{abstract}


\maketitle

\section{\label{sec:intro}Introduction}
Repetitive pacing of biological reaction diffusion media by over-threshold stimuli gives rise to periodically propagating excitation waves. In cardiac tissue the duration of excitation referred to as action potential duration, $T_{AP}$, as well as the speeds of excitation wavefront and waveback, depend on previous stimulation periods and refractory (diastolic) intervals, $T_{DI}$ \cite{chialvo1990, gilmour1997, cain2004}. The analysis of such dependences known as restitution curves has been established as an effective method for evaluation of normal functioning of the heart \cite{karma1994, fenton1998, garfinkel2000, qu2000, schwartz2000, chernyak1999}.

Pioneering experimental studies \cite{boyett1978, elharrar1983, franz1988} established two major pacing sequences such as steady state (dynamic) and $\text{S}1\text{-S}2$ pacing protocols, which led to two standard $T_{AP}$ restitution dependences. It has been shown that the protocol dependent rates of $T_{AP}$ adaptation were different for stepwise $\text{S}1\text{-S}2$ perturbations of cycle length acceleration and deceleration \cite{franz1988, arnold1982}. This phenomenon was later experimentally introduced as action potential duration cycle length hysteresis \cite{sarma1987, krahn1997} and has been recently associated with cardiac ischemia, coronary flow reduction \cite{lauer2006, starobin2007}, and cardiac memory \cite{berger2004, wu2004}.

A stepwise change in pacing rate following a long series of conditioning S1 stimuli may result in prolonged adaptation of action potential duration to its new steady-state value. The process of such adaptation can extend well beyond the $T_{AP}$ response to the first S2 test stimulus \cite{elharrar1983, franz1988} and give rise to the additional constant BCL (basic cycle length) transient restitution component attributed to cardiac memory \cite{gilmour1997, hall1999, otani1997, kalb2004}. Relationship between this phenomenon and stability of pulse propagation as well as with rate dependent $T_{AP} \text{-} T_{DI}$ hysteresis has been recently investigated in experimental \cite{starobin2007, hall1999, kalb2004, watanabe2002} and theoretical \cite{fenton1999, cherry2004, tolkacheva2002, tolkacheva2003, kalb2005} studies.

It has been found that in the presence of restitution transients cardiac dynamics is more complex than predicted by Nolasco and Dahlen restitution criterion \cite{nolasco1968}. Specifically, it was demonstrated that dynamic restitution curve slopes $> 1$ may not automatically indicate loss of excitation wave stability and subsequent appearance of alternans \cite{hall1999, fenton1999, cherry2004, tolkacheva2002, tolkacheva2003, banville2002}. Computational experiments with Fenton-Karma and Mitchell-Schaeffer ionic models readily identified that such effects can be quantified by fitting $T_{AP}$ and $T_{DI}$ for specific restitution curves using time dependent gate variables \cite{fenton1999, cherry2004, mitchell2003, schaeffer2007, shiferaw2005}.

In this paper, we analyze excitation wave propagation in a one-dimensional cable based on the approach, which follows from direct experimental observations of the dependence of cardiac muscle resting potential on frequency of external pacing \cite{attwell1981, davidenko1990, ravens1998}. We implement a two-variable exactly solvable Chernyak-Starobin-Cohen (CSC) reaction-diffusion model \cite{chernyak1998a, chernyak1998b} and modify it accordingly to incorporate pacing rate driven adjustments of resting potential as a rate dependent excitation threshold. We use an exponential-like evolution of excitation threshold, $V_r$, that takes place over the course of multiple heart beats following stepwise changes in pacing rate \cite{davidenko1990}. We demonstrate that adaptation of $V_r$ significantly affects the stability of pulse propagation and gives rise to rate dependent $T_{AP} \text{-} T_{DI}$ hysteresis. In particular, we find that under given medium parameters, regardless of the slopes of restitution curves, the appearance of alternans is determined by proximity of the wavefront speed to the minimal speed of a stationary solitary pulse determined analytically in \cite{chernyak1998a}.

\section{\label{sec:methods}Methods}
Basic equations that describe a class of exactly solvable models for excitable media have been defined in \cite{chernyak1998a, chernyak1998b}. Here we introduce a modification of this analytical model by adjusting the excitation threshold, $V_r$, in response to changes in frequency of external pacing. We will consider the model in dimensionless form:
\begin{eqnarray}
\frac{\partial u}{\partial t} &=& \frac{\partial^2u}{\partial x^2} - i (u, v) + P (x, t) \label{eq:one} \\
i(u, v) &=& 
\begin{cases}
\lambda u & \text{for $u < v$}\\
(u - 1) & \text{for $u \ge v$}
\end{cases} \nonumber \\
\frac{\partial v}{\partial t} &=& \epsilon(\zeta u + V_r - v) \label{eq:two} \\
\frac{d V_r}{dt} &=& \frac{-V_r+B(t)}{\tau} \label{eq:three}
\end{eqnarray}
$u(x, t)$ and $v(x, t)$ are a membrane potential and slow recovery current, respectively. $\lambda$, $\epsilon$, $\zeta$, and $\tau$ are the model parameters, where $\tau^{-1} < \epsilon$. The scaling of the system is described in the Appendix.

The pacing function, $P (x, t)$, is defined as a product of two functions, $X(x)$ and $T(t)$. Each is composed of the Heaviside step function, $\Theta (x)$, as follows: $X(x) = A[\Theta (x - \delta_1) - \Theta (x - \delta_2)]$ and $T (t) = \Theta (t_k) - \Theta (t_k + T_s)$, where $A$ and $\delta_2 - \delta_1$ are the amplitude and width of the pulse, respectively, $T_s$ is the pulse duration, and $t_k = t_{N(m-1)} + [k - N (m - 1)]T_m$ are the instants of time when stimuli are delivered. We use this to construct a pacing protocol where the pacing period, $T_m$, is stepwise constant ($T_m > T_s$). $N$ represents the number of stimuli at each pacing interval plateau, the index $m$ denotes each pacing plateau, $m = 1, \ldots , M$, and $k$ is an integer in the range $k = 1, \ldots , N M$. The number of stimuli $N$ is the same for all plateaus.  Overall during the course of the protocol, $T_m$ progressively decreases to a minimum and then increases to its starting value.

The right-hand term $B$ in Eq.~\ref{eq:three} responds to the stepwise evolution of pacing period, $T_m$.  Stepwise changes in $B$ result in smooth exponential transition of the excitation threshold from one steady-state plateau to another.  A steady-state value of excitation threshold at each $m^{\text{th}}$ plateau, $V_r^m \equiv B_m^{\pm}$, was chosen to be linearly dependent on the corresponding pacing interval, $T_m$ \cite{attwell1981, davidenko1990}.
\begin{equation}
B_m^{\pm} = -\beta^{\pm} T_m + \alpha^{\pm} \label{eq:four}
\end{equation}
Here $\alpha^{\pm}$ and $\beta^{\pm}$ are positive parameters that determine the amplitude of change of $B_m^{\pm}$ between two consecutive pacing plateaus for the increasing and decreasing rate, respectively.

\section{\label{sec:numerics}Numerical Simulations}
\begin{figure*}
\includegraphics{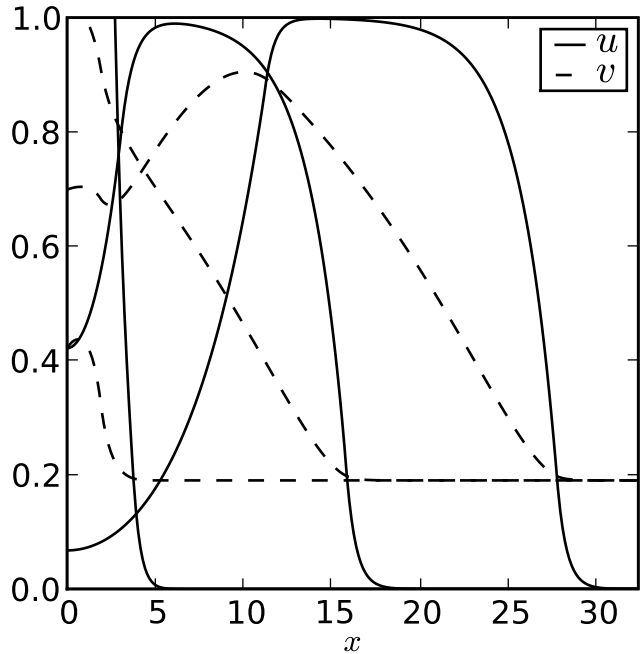}
\caption{\label{fig:figone}Spatio-temporal dynamics of $u$,$v$ variables. Three snapshots of $u$ and $v$ are shown for a excitation threshold $V_r = 0.19$. The first snapshot occurs at the instant of the pacing stimulus and shows the formation of the wavefront. The next two snapshots are taken after periods of $7.2$ ($20\%$ of the pacing period) and $14.4$ and show formation and propagation of the pulse.}
\end{figure*}

The system of Eqs.~\ref{eq:one}-\ref{eq:three} was solved numerically on a short cable of $250$ grid points with spatial and temporal grid intervals of $\Delta x = 0.13$ and $\Delta t = 7.2 \times 10^{-4}$, respectively. The length of the cable was approximately equal to the width of the pulse to reflect the relative dimensions of the heart and a propagating cardiac pulse at moderate heart rates. Periodic wavetrains were produced by stimulating the cable with a square wave at the left end using the function $P (x, t) = X(x)T (t)$, defined above, where $A = 10$, $\delta_1 = 2\Delta x$, $\delta_2 = 15\Delta x$, and $T_s = 10^3\Delta t$. The model parameters $\lambda$, $\epsilon$, and $\zeta$ were equal to $0.4$, $0.1$, and $1.2$, respectively, for all simulations. Numerical solutions were computed using a second-order explicit-difference scheme \cite{chernyak1999}. A typical solution is depicted in Fig.~\ref{fig:figone}, showing the propagation of a single pulse at three instants of time.

In order to quantify the dynamics of the system (Eqs.~\ref{eq:one}-\ref{eq:three}), we computed the action potential duration, $T_{AP}$, the diastolic interval, $T_{DI}$, and the wavefront velocity, $c$. The action potential duration was defined as the interval of time when $u > v$ at a specified node, $x_0$. Accordingly, the refractory period, $T_{DI}$, was defined as the interval of time when $u < v$. The speed of the wavefront, $c$, was calculated based on the time it took for a point of constant phase on the wavefront ($u = 0.5$) to travel a span of $10$ grid points centered around $x_0$. In order to analyze a developed pulse, we measured intervals and speeds at $x_0 = 20$ (except Sec.~\ref{sec:propagationD}) where the speed of the wavefront had reached a constant steady-state value.

\subsection{\label{sec:constantvrA}Restitution for constant excitation threshold} 
The system of equations~\ref{eq:one}-\ref{eq:three} was initially studied with Eq.~\ref{eq:three} replaced by its asymptotic form $\frac{\partial V_r}{\partial t} = 0$, which is equivalent to $\tau = \infty$. The cable was stimulated periodically for forty consecutive pacing periods over the range $T_m = 70$ to $25$ with decrements of $1.5$.  At the end of each plateau, $T_{AP}$ and $T_{DI}$ had reached steady-state values that were used to compose the steady-state restitution curve.  Two values of $T_m$, at $30$ and $27$, were used to obtain conventional $\text{S}1\text{-S}2$ restitution curves\cite{boyett1978, elharrar1983}.  The pulse, $T^{n+1}_{AP}$, resulting from the test stimulus, S$2$, following the conditioning sequence and the last diastolic interval, $T^n_{DI}$, from the conditioning S$1$ plateau composed the $\text{S}1\text{-S}2$ restitution curve.

\begin{figure*}
\includegraphics{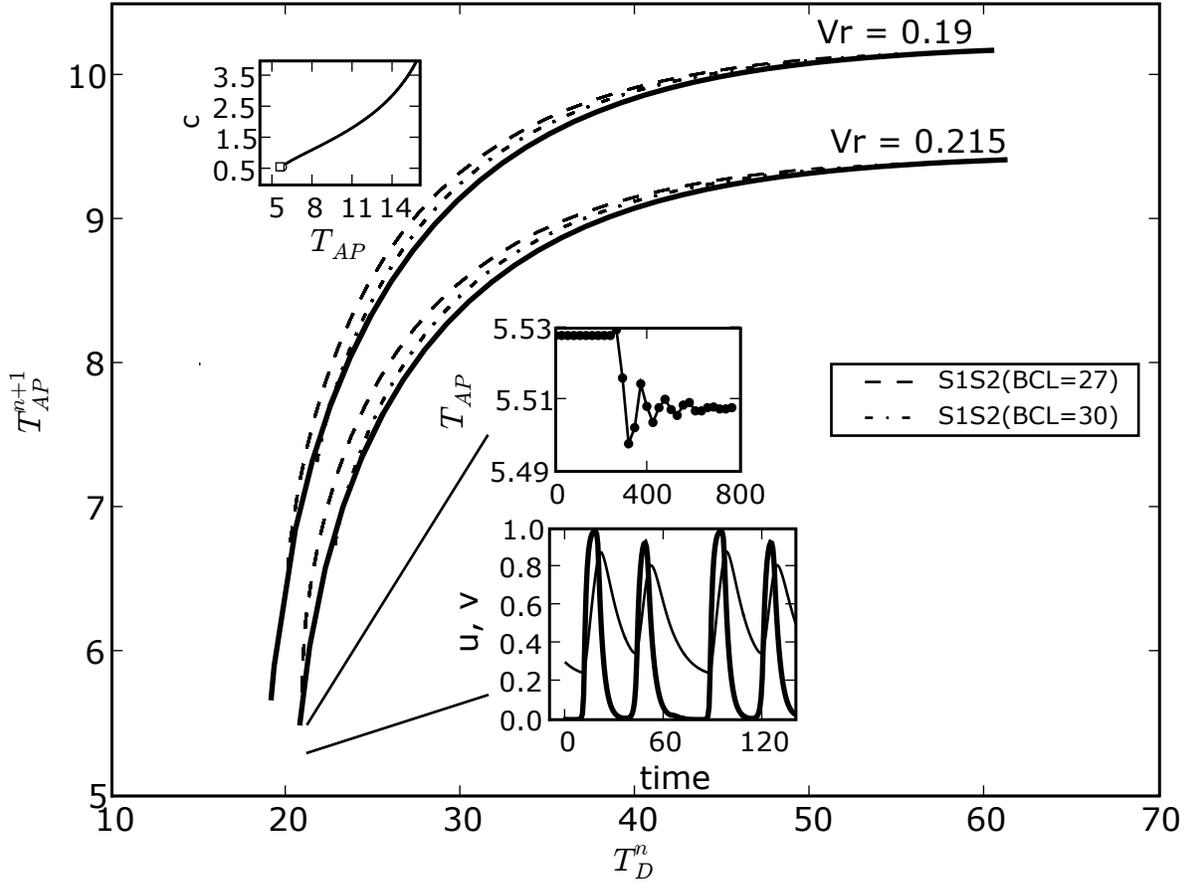}
\caption{\label{fig:figtwo}Steady-state restitution curves are shown with solid lines, and $\text{S}1\text{-S}2$ restitution curves computed for $\text{S}1$ BCLs of $27$ and $30$ are shown with dashed lines. Critical $T_{AP}$ and $T_{DI}$ values at which pulse durations begin to oscillate are at the ends of the curves where $T^{crit}_{DI} = 19.1$, $T^{crit}_{AP} = 5.7$ and $T^{crit}_{DI} = 20.8$, $T^{crit}_{AP} = 5.5$ for $V_r = 0.19$ and $V_r = 0.215$, respectively.  Open markers in the upper left insert indicate corresponding critical speeds plotted with the analytical dispersion curve for a solitary pulse \cite{chernyak1998a}.  The mid and lower inserts demonstrate oscillation of pulse duration at and below the critical speed.}
\end{figure*}

Steady state and $\text{S}1\text{-S}2$ restitution curves computed for a pair of constant excitation thresholds and a pair of $\text{S}1$ basic cycle lengths are shown in Fig.~\ref{fig:figtwo}. We observed that the maximal difference between steady state and $\text{S}1\text{-S}2$ restitution curves comprised less than $15\%$ of the corresponding steady state value of $T_{AP}$, which implied that transient responses to any premature stimulus were on average limited to just a single non-stationary pulse.

The ends of both steady state restitution curves shown in Fig.~\ref{fig:figtwo} indicate the critical points below which no stable propagation and no $1 : 1$ responses were observed. We found that when the pacing interval reached the value of $T_m = 26.3$ at $V_r = 0.215$, the resulting action potentials oscillated in duration as shown in the mid level insert panel. Further reduction of the pacing interval from $T_m = 26.3$ to $T_m = 25.9$ induced an even more complex $3 : 2~T_{AP}$ response pattern depicted in the lower level insert panel.

The upper level insert shows the dispersion curve computed analytically for a steady state solitary pulse \cite{chernyak1998a}. The critical speed and duration of a stable solitary pulse for the model parameters, $\lambda$, $\zeta$, and $\epsilon$, described above is $c_{crit}=0.48$ and $T^{crit}_{AP}=5.4$.  As shown in the insert, the difference between the numerical and analytical critical speeds is relatively small and amounts only to $10\%$ of the wavefront speed determined from our numerical model at $V_r = 0.215$. This suggests a criterion for determining the critical speed below which pulse durations start to oscillate.  On the contrary, the conventional Nolasco-Dahlen critical slope stability criterion is unsuitable as the maximum slope of the dynamic restitution curve at this $V_r$ is $30\%$ greater than one \cite{nolasco1968}.

\subsection{\label{sec:ratedepvrB}Restitution and hysteresis for rate-dependent $V_r$} 
\begin{figure*}
\includegraphics{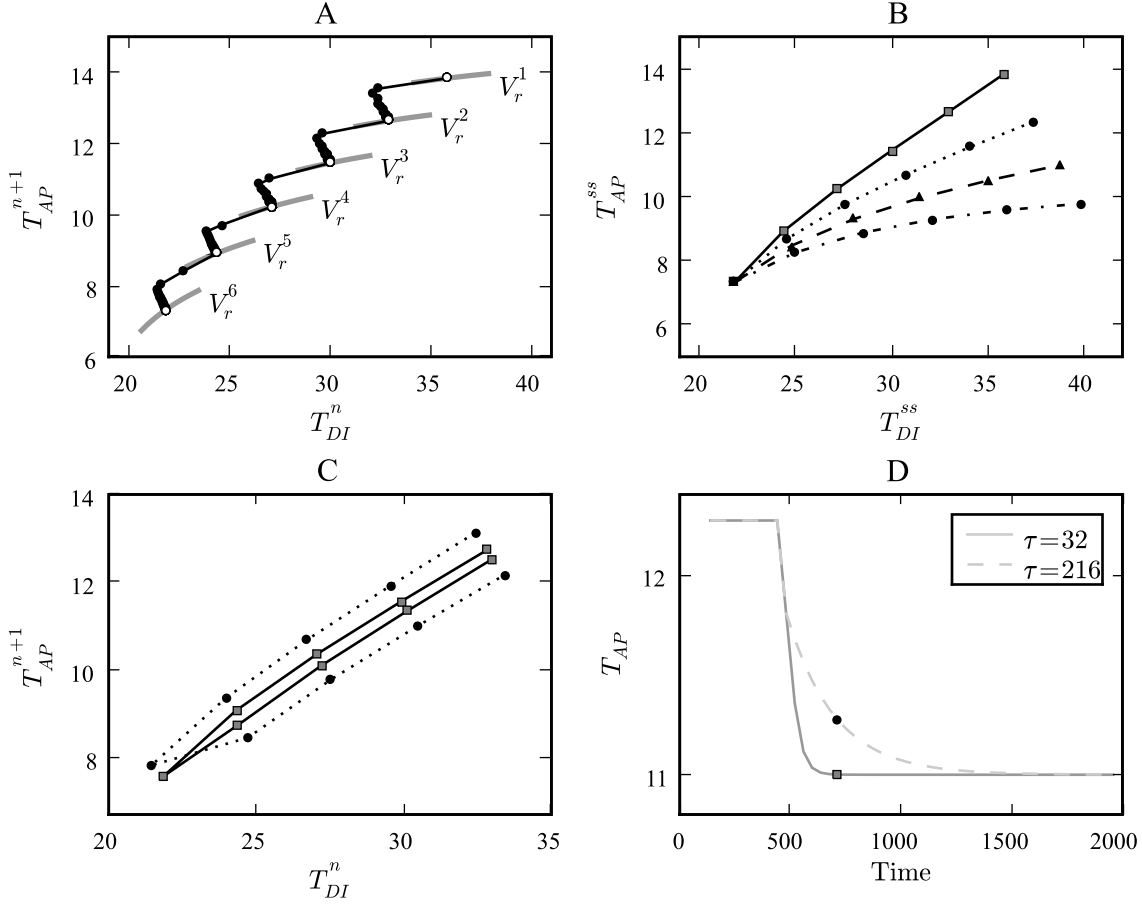}
\caption{\label{fig:figthree}Restitution relations and $T_{AP} \text{-} T_{DI}$ hysteresis for a stepwise stimulation protocol with rate dependent excitation threshold. (A) A series of five stepwise changes of pacing rate are shown with individual $T_{AP}$ and $T_{DI}$ as black circles. Open circles mark steady-state $T_{AP}$ and $T_{DI}$ intervals. Portions of constant excitation threshold, $V^m_r \equiv B_m^{\pm}$, and steady state restitution curves are shown in gray. (B) $T_{AP} \text{-} T_{DI}$ hysteresis produced by increasing differences between $B_m^{+}$ and $B_m^{-}$. $\beta^{+} = 6 \times 10^{-3}$ and $\alpha^{+} = 0.37$ for the top curve (solid line, squares), $\beta^{-} = 4 \times 10^{-3}$ and $\alpha^{-} = 0.31$ (dotted line, solid circles), $\beta^{-} = 2 \times 10^{-3}$ and $\alpha^{-} = 0.25$ (dashed line, triangles), and $\beta^{-} = 0.1 \times 10^{-3}$ and $\alpha^{-} = 0.19$ for the lowest curve (dash-dotted line, circles). (C) $T_{AP} \text{-} T_{DI}$ hysteresis for different values of the adaptation constant, $\tau$ ($\tau = 216$, dotted line; $\tau = 32$, solid line), when $\beta^{+} = \beta^{-}$ and $\alpha^{+} = \alpha^{-}$. (D) Adaptation of action potential duration after an abrupt decrease in stimulation interval for two different rate constants ($\beta^{+} = 6 \times 10^{-3}$ and $\alpha^{+} = 0.37$). The measured transient intervals in Panel C are highlighted with corresponding markers.}
\end{figure*}

Unlike the previous section, the evolution of $V_r$ according to Eq.~\ref{eq:three} resulted in a set of prolonged transients initiated by abrupt changes in stimulation rate. These $T_{AP}$ transients, which constituted the constant BCL restitution (negative slope, Fig.~\ref{fig:figthree}A), had a duration of $5$-$50$ stimulation periods depending on the adaptation constant, $\tau$ (Fig.~\ref{fig:figthree}D). The phase of constant BCL adaptation followed the immediate $\text{S}1\text{-S}2$ responses, which were on the contrary positive and aligned with the steady state restitution curves for each steady state excitation threshold (gray lines, Fig.~\ref{fig:figthree}A).

Different steady state values of excitation threshold during progressively increasing and decreasing stimulation rates gave rise to $T_{AP} \text{-} T_{DI}$ hysteresis. The cable was stimulated periodically for fifty consecutive periods at a series of $\text{S}1$ conditioning plateaus. The series consisted of five plateaus with decreasing $T_m$ followed by the same number of plateaus with increasing $T_m$. The excitation threshold evolved according to Eq.~\ref{eq:three}, and its steady state value at each plateau was related to the pacing period by Eq.~\ref{eq:four}.

Using this protocol we demonstrated that higher values of $T_{AP} \text{-} T_{DI}$ hysteresis corresponded to greater differences between $B_m^{+}$ and $B_m^{-}$ during the decremental and incremental stages of the pacing protocol (Fig.~\ref{fig:figthree}B). We also found that $T_{AP} \text{-} T_{DI}$ hysteresis was dependent on the adaptation constant, $\tau$, if $T_{AP}$ and $T_{DI}$ were measured before their steady-state values were reached. For instance, for a time lag of seven stimulation intervals (Fig.~\ref{fig:figthree}D), the magnitude of $T_{AP} \text{-} T_{DI}$ hysteresis increased as the adaptation constant increased from $\tau = 32$ to $\tau = 216$ (Fig.~\ref{fig:figthree}C).

\subsection{\label{sec:alternansC}Transitions to $T_{AP}$ alternans, and comparison with stability of a solitary pulse} 
When the pulse duration and speed are less than certain critical values, steady-state solitary pulse and wavetrain propagation in an infinite cable do not exist \cite{chernyak1998a, chernyak1998b}. Such critical values for a solitary pulse occur at the end of the solitary pulse's dispersion curve, as computed analytically in \cite{chernyak1998a}. Similar critical values for our short cable were described earlier in Sec.~\ref{sec:constantvrA} for a medium with constant excitation threshold.  In this section, we analyze perturbations of $T_{AP}$ and wavefront speed when steady state pulse speeds are below a certain value close to the critical speed of a solitary pulse.

\begin{figure*}
\includegraphics{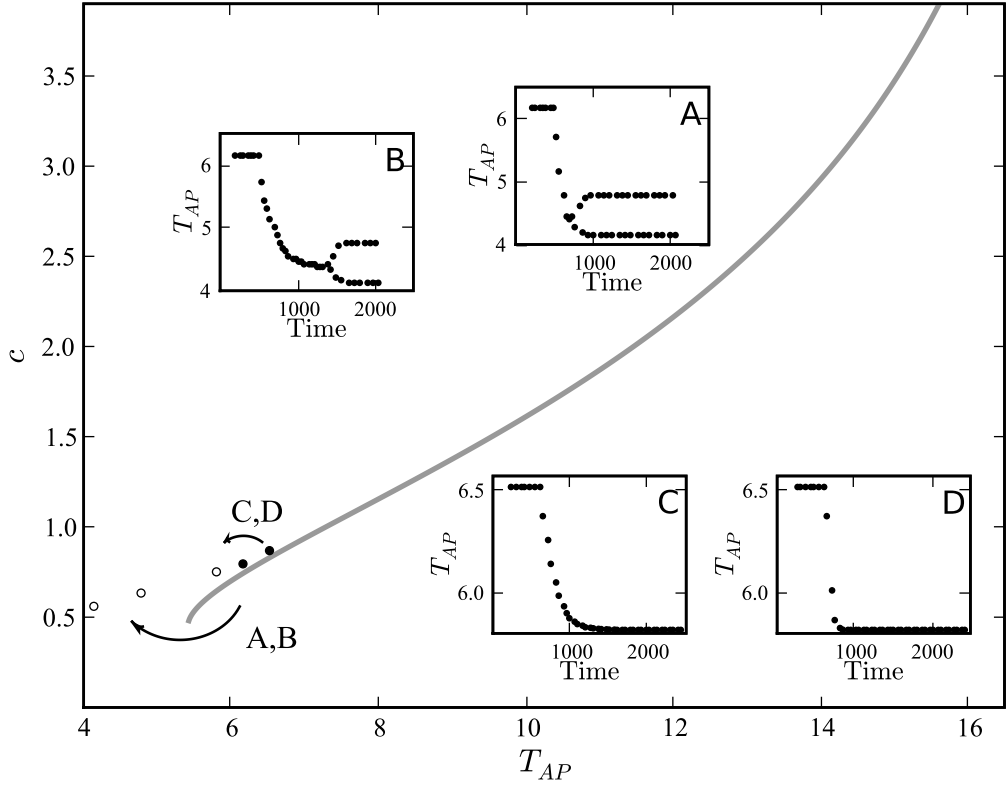}
\caption{\label{fig:figfour}Solitary pulse dispersion curve and initiation of $T_{AP}$ alternans at $x_0 = 20$. Inserts (A) and (B) show $T_{AP}$ response to a perturbation in stimulation rate, for initial $T_m = 46.8$, final $T_m = 40.3$, initial $B_m^{+} = 0.31$, and final $B_m^{+} =0.32$. The adaptation constant is small, $\tau = 32$, in A and large, $\tau = 216$, in B. Inserts (C) and (D) depict stable transitions of $T_{AP}$ for the same adaptation constants ($\tau = 216$ in C; $\tau = 32$ in D) when the starting and ending pulse speeds are greater than in (A) and (B) (initial $T_m = 51.3$, final $T_m = 45.0$, initial $B_m^{+} = 0.31$, and final $B_m^{+} =0.32$). Gray line in main panel depicts dispersion curve of a steady-state solitary pulse \cite{chernyak1998a}. The markers at the left end of the dispersion curve show the starting speed and $T_{AP}$ (solid circles) and the ending speeds and $T_{AP}$ (open circles) for both inserts.}
\end{figure*}

\begin{figure*}
\includegraphics{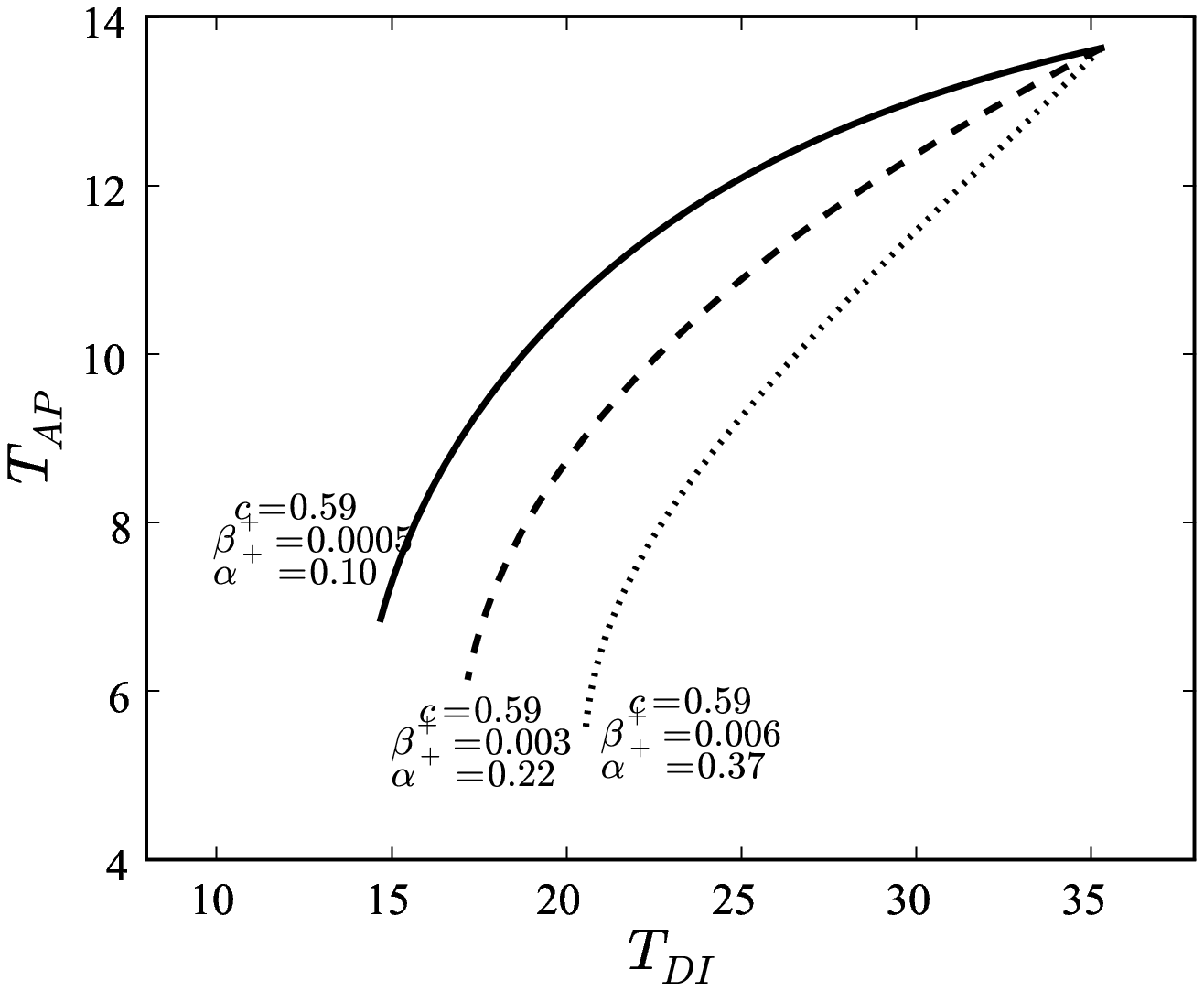}
\caption{\label{fig:figfive}Steady state restitution for three different sets of rate dependence parameters, $\beta^+$ and $\alpha^{+}$. $\tau = 216$ for all curves.  $x_0 = 20$.  The ends of the curves indicate the values of $T_{AP}$ below which the duration of pulses oscillates.  Note that the speed of the pulses at the ends of the curves is the same regardless of the magnitudes of restitution slopes, which vary with parameters $\beta^+$ and $\alpha^{+}$.}
\end{figure*}

We perturbed both pulse duration and wavefront speed near their analytical critical values, $T^{crit}_{AP}$ and $c_{crit}$, as indicated by arrows labeled ``A,B'' and ``C,D'' (Fig.~\ref{fig:figfour}). If the speeds of the wavefronts elicited after perturbation exceeded $c_{crit}$  less than $22\%$, we observed $T_{AP}$ alternans (inserts A and B) that developed between two values indicated by open markers at the end of the arrow ``A,B''.  Alternans occurred sooner when the adaptation rate was higher due to a smaller adaptation constant, $\tau$ (insert A). The minimal non-alternating wavefront speeds computed at $x_0 = 20$ for different $B_m^{+}$ were the same regardless of changes of $B_m^{+}$ or the four fold increase of the steady-state restitution slope (Fig.~\ref{fig:figfive}).  On the contrary, when the perturbation resulted in a wavefront speed that exceeded $c_{crit}$ by more than $22\%$ (the end of the arrow ``C,D''), similar bifurcations did not occur. Instead, we observed typical exponential adaptation from one steady-state to another (inserts C and D).  

\subsection{\label{sec:propagationD}Influence of propagation on the spatial distribution of hysteresis and alternans}
\begin{figure*}
\includegraphics{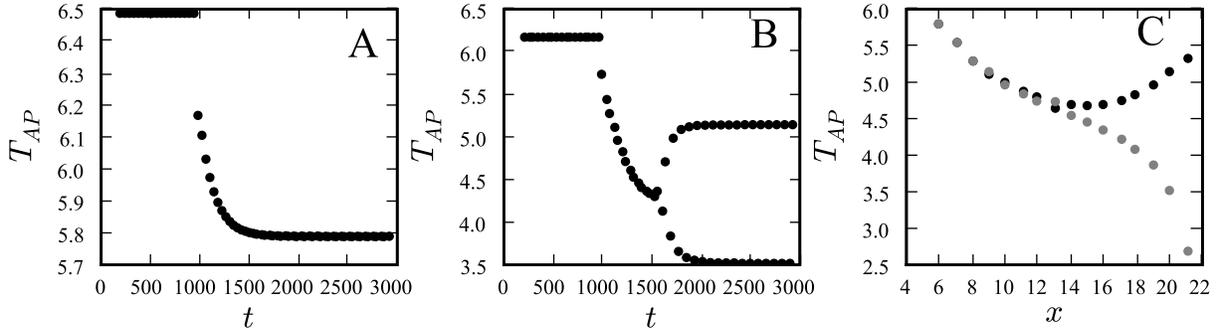}
\caption{\label{fig:figsix}The spatial distribution of $1:1$ and $2:2$ responses along the cable. Oscillations were induced by perturbing $T_{AP}$, with initial $T_m = 46.8$, final $T_m=40.2$, initial $B_m=0.31$, and final $B_m=0.32$. Panels A and B show the response to perturbation at $x_0 = 6$ and $x_0 = 20$, respectively.  Panel C shows the spatial distribution of the last two $T_{AP}$ of the train of $50$ pulses illustrated in Panels A and B.}
\end{figure*}

We demonstrate that $T_{AP}$ adaptation to a stepwise change in pacing interval is different for different points of observation along the cable.  When the distance between the point of observation and stimulation site increases, the wavefront and waveback speeds decrease.  At the point $x_0=6$ adjacent to the stimulation site, for an initial $T_m=46.8$ and final $T_m=40.2$, the wavefront speed, $c=0.69$, is substantially higher than the critical speed of a solitary pulse, $c_{crit}=0.48$.  Under these conditions, $T_{AP}$ gradually adapts to a new steady state value as shown in Fig.~\ref{fig:figsix}A.  On the contrary, at $x_0 = 20$, for the same change in pacing interval and excitation threshold parameter, the wavefront speed is substantially lower, $c=0.59$, which results in a series of oscillating $T_{AP}$.  The lower branch of alternating $T_{AP}$ (Fig.~\ref{fig:figsix}B) corresponds to slowly propagating wavefronts whose speeds are virtually equal ($7\%$ difference) to the analytical value of $c_{crit}$.  Further increase of the distance between the observation point and stimulation site results in the increase of the amplitude of $T_{AP}$ alternans.  The closest point at which alternans can be observed is located at the midpoint of the cable (Fig.~\ref{fig:figsix}C).

\begin{figure*}
\includegraphics{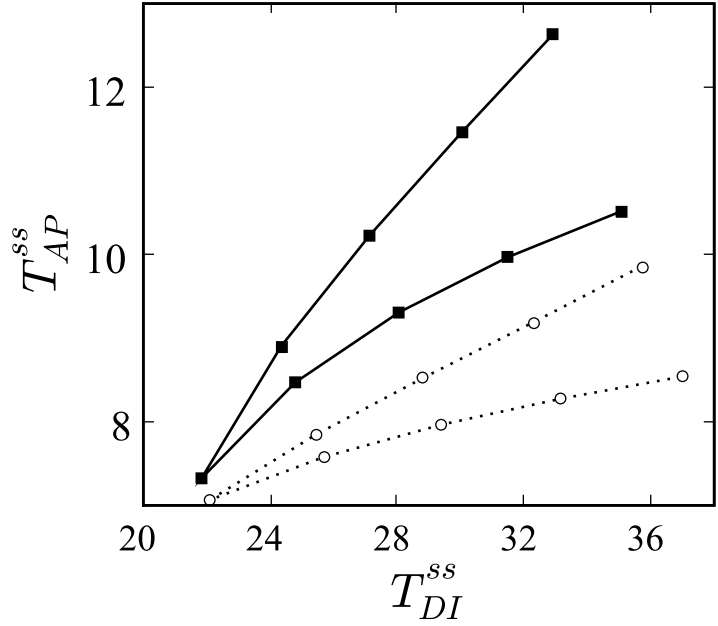}
\caption{\label{fig:figseven}Steady-state $T_{AP} \text{-} T_{DI}$ hysteresis at $x_0 = 4$ (dashed line, open circles) and $x_0 = 20$ (solid line, squares). For both curves, $\beta^{+} = 6 \times 10^{-3}$, $\alpha^{+}=0.37$, $\beta^{-} = 2 \times 10^{-3}$, and $\alpha^{-}=0.25$.}
\end{figure*}

We also observed that the magnitude of $T_{AP} \text{-} T_{DI}$ hysteresis is larger when measured further from the stimulation site.  Figure~\ref{fig:figseven} shows the twofold increase of the magnitude of hysteresis and $T_{AP}$ between two observation points at $x_0=4$ and $x_0=20$.

\section{\label{sec:conclusions}Conclusions}
We have demonstrated that rate dependent restitution and $T_{AP} \text{-} T_{DI}$ interval hysteresis can be reproduced in a one-dimensional two-variable CSC reaction-diffusion medium where excitation threshold adjusts to changes in pacing rate. We show that the rate dependence of restitution and hysteresis are influenced by two major factors. The first one is the adaptation constant, $\tau$, of excitation threshold evolution after an abrupt change in pacing interval.  The second is the dependence, $B_m^{\pm}$, of the steady-state excitation threshold on the pacing period.  We show that steady-state and $\text{S}1\text{-S}2$ restitution curves coincide if the excitation threshold is constant, which corroborates with earlier findings for other reaction diffusion models \cite{cain2004, tolkacheva2002, kalb2005}. On the contrary, the steady-state and $\text{S}1\text{-S}2$ restitution curves diverge if the magnitude of excitation threshold varies with changes in pacing rate, leading to prolonged $T_{AP}$ transients following an abrupt change in pacing interval.

We found that larger values of $T_{AP} \text{-} T_{DI}$ interval hysteresis were associated with greater differences between $B_m^{+}$ and $B_m^{-}$. Even if there was no difference between these values, hysteresis resulted from increasing adaptation constants, $\tau$, if $T_{AP}$ and $T_{DI}$ were measured before their steady-state values were reached.

Our numerical simulations showed that stimulating the cable with short pacing intervals and high excitation thresholds elicited slower pulses that led to $T_{AP}$ alternans.  It was observed that the minimal stable wavefront speed could be approximated by the analytical critical speed of a stable solitary pulse \cite{chernyak1998a, chernyak1998b}. This approximation was suitably accurate regardless of values of $B_m^{\pm}$ and magnitudes of the slopes of restitution curves.  We also found that the onset of alternans occurring after an abrupt change in pacing rate was more delayed for larger values of $\tau$.

\appendix*

\section{\label{app:appen}}
The scale of $u$ is the maximum steady-state action potential amplitude $U_0$, the scale of $v$ is given by $\sigma_f U_0$, and the time scale is $C_m/\sigma_f$, where $\sigma_f$ corresponds to the maximum sodium conductance and $C_m$ is the membrane capacitance. The characteristic length scale is given by $\sqrt{D/\sigma_f}$, where $D$ is the diffusion coefficient. The small parameter $\epsilon \ll 1$, is equal to $C_m/(\tau_s\sigma_f )$ and $\zeta = \sigma_s/\sigma_f$ where $\sigma_s$ corresponds to the maximum potassium conductance.

\begin{acknowledgments}
This research was funded by Mediwave Star Technology, Inc. and was partially supported by the University of North Carolina at Greensboro. We are grateful to Lanty L. Smith and Thomas R. Sloan for their continuous support.  We would also like to thank David Schaeffer, Wanda Krassowska, and Daniel Gauthier for helpful discussions and critical reviews.
\end{acknowledgments}

\bibliography{bib_master_copy}

\end{document}